\documentclass[manuscript]{aastex}

\usepackage{amsmath}
\usepackage{amssymb}
\usepackage{graphicx}
\usepackage{color}
\usepackage{epstopdf}

\def\gtaprx {\lower .1ex\hbox{\rlap{\raise .6ex\hbox{\hskip .3ex
    {\ifmmode{\scriptscriptstyle >}\else
        {$\scriptscriptstyle >$}\fi}}}
    \kern -.4ex{\ifmmode{\scriptscriptstyle \sim}\else
        {$\scriptscriptstyle\sim$}\fi}}}
\def\ltaprx {\lower .1ex\hbox{\rlap{\raise .6ex\hbox{\hskip .3ex
    {\ifmmode{\scriptscriptstyle <}\else
        {$\scriptscriptstyle <$}\fi}}}
    \kern -.4ex{\ifmmode{\scriptscriptstyle \sim}\else
        {$\scriptscriptstyle\sim$}\fi}}}
\newcommand{\note}[1]{\emph{\textcolor{black}{}}}

\newcommand{\Ms}{{\ensuremath{{M}_{\odot} }}}

\newcommand{\Ni}{{\ensuremath{^{56}\mathrm{Ni}}}}

\newcommand{\He}{{\ensuremath{^{4} \mathrm{He}}}}

\newcommand{\Ox}{{\ensuremath{^{16}\mathrm{O}}}}

\newcommand{\Si}{{\ensuremath{^{28}\mathrm{Si}}}}
\newcommand{\Mg}{{\ensuremath{^{24}\mathrm{Mg}}}}
\newcommand{\Cx}{{\ensuremath{^{12}\mathrm{C}}}}

\newcommand{\cm}{{\ensuremath{\mathrm{cm}}}}
\newcommand{\erg}{{\ensuremath{\mathrm{erg}}}}

\newcommand{\CASTRO}{\texttt{CASTRO}}
\newcommand{\KEPLER}{\texttt{KEPLER}}
\newcommand{\STELLA}{\texttt{STELLA}}
\newcommand{\RAGE}{\texttt{RAGE}}

\newcommand{\gcc}{\ensuremath{\mathrm{g}\,\mathrm{cm}^{-3}}}

\makeatletter

\newcommand{\Rmnum}[1]{\expandafter\@slowromancap\romannumeral #1@}
\makeatother

\begin{document}
\title{Two-Dimensional Simulations of Pulsational Pair-Instability Supernovae} 

\author{ Ke-Jung Chen\altaffilmark{1,2,*}, Stan Woosley\altaffilmark{1}, 
Alexander Heger\altaffilmark{3}, Ann Almgren\altaffilmark{4}  and 
Daniel J. Whalen\altaffilmark{5,6}} 

\altaffiltext{*}{IAU Gruber Fellow; kchen@ucolick.org} 

\altaffiltext{1}{Department of Astronomy \& Astrophysics, University of California, Santa 
Cruz, CA 95064, USA} 

\altaffiltext{2}{School of Physics and Astronomy, University of Minnesota, Minneapolis, MN 
55455, USA}

\altaffiltext{3}{Monash Centre for Astrophysics, Monash University, Victoria 3800, Australia} 

\altaffiltext{4}{Center for Computational Sciences and Engineering, Lawrence Berkeley 
National Lab, Berkeley, CA 94720, USA}

\altaffiltext{5}{T-2, Los Alamos National Laboratory, Los Alamos, NM 87545, USA} 

\altaffiltext{6}{Zentrum f\"{u}r Astronomie, Institut f\"{u}r Theoretische Astrophysik, 
Universit\"{a}t Heidelberg, Albert-Ueberle-Str. 2, 69120 Heidelberg, Germany}

\begin{abstract} 

Massive stars that end their lives with helium cores in the range of 35 to 65 \Ms\ are known 
to produce repeated thermonuclear outbursts due to a recurring pair-instability.  In some 
of these events, solar masses of material are ejected in repeated outbursts of several times 
10$^{50}$ erg each. Collisions between these shells can sometimes produce very luminous 
transients that are visible from the edge of the observable universe.  Previous 1D studies of 
these events produce thin, high-density shells as one ejection plows into another.  Here, in 
the first multi-dimensional simulations of these collisions, we show that the development of a 
Rayleigh-Taylor instability truncates the growth of the high density spike and drives mixing
between the shells.  The progenitor is a 110 \Ms\ solar-metallicity star that was shown in 
earlier work to produce a superluminous supernova.  The light curve of this more realistic 
model has a peak luminosity and duration that are similar to those of 1D models but a 
structure that is smoother.

\end{abstract}

  
\section{Introduction}

The idea of a ``pulsational pair-instability supernova'' (PPISN) was introduced by 
\citet{barkat1967} and explored in some detail by \citet{woosley2007}. For a range of helium 
core masses above approximately 35 \Ms, the production of electron-positron pairs 
{\color{black} occurs during central neon burning} and triggers an instability that leads to rapid 
contraction of the core and explosive nuclear burning.  If the helium core mass is above 
about 65 \Ms, {\color{black} the pair production instability occurs after carbon ignition} and the 
energy released can completely unbind the star in a single explosive event known as a 
pair-instability supernova \citep[PISN;][]{heger2002,scannapieco2005,chat2012b, chat2013, chen2014b}.  
If the core mass is above 133 \Ms, for reasonable nuclear reaction rates, convection theory, 
and no rotation, nuclear burning is unable to reverse the collapse and a black hole forms 
\citep{heger2002}.  

For helium core masses from 35 - 65 \Ms,  Although explosive burning is violent and 
energetic, it cannot unbind the entire star, and cycles of instability and mass ejection 
can instead occur.  
The energy, duration, and mass ejected by these nuclear-powered pulses increases as the 
helium core mass rises, and by about 45 \Ms\ they are sufficient to produce SN-like displays {\color{black} \citep{woosley2007,vink2014}}. 
If the helium core is capped by a substantial hydrogen envelope, the first strong pulse ejects 
it.  Depending on the radius and mass of the envelope, this initial ejection may give rise to 
either a faint or rather typical Type IIp SN. Subsequent pulses later overtake and collide with 
the first and produce a much brighter Type IIn SN.  If there is no hydrogen envelope, 
collisions between helium shells can produce a bright Type I SN.  The helium core mass 
range from 45 to 55 \Ms\ is particularly interesting because the characteristic time scale 
between pulsations is years and the collisions between shells ejected with speeds $\sim$ 
1000 km s$^{-1}$ occur at 10$^{15}$ - 10$^{16}$ cm, where the collision energy is mostly 
dissipated by optical emission.  Since the collision energy can approach 10$^{51}$ erg, a 
superluminous event can result.

The discovery of both PI and PPI SN candidates in the local universe, and the realization 
that such events might be visible at high redshifts, have excited interest in this exotic 
explosion mechanism.  SN 2007bi at $z =$ 0.127 \citep{2007bi} and SN 2213-1745 at $z 
=$ 2.06 \citep{cooke12} are PISN candidates, and SN 1000$+$0216 at $z =$ 3.90 \citep{
cooke12} and perhaps SN 2006oz at $z =$ 0.376 \citep{lel12} may be PPISNe.  {\color{black} 
\citet{dessart2013} suggest that SN 2007bi may be a magnetar spin-down powered event
\citep{kasen2010, woosley2010, dessart2012} or it could be due to an interaction between 
H-poor SN ejecta and a circumstellar medium \citep{chat2012a, moriya2013}. The nature 
of these transients is still under debate.} \citet{wet13d} have now shown that PPISNe, like 
PISNe and Type IIne, may be visible to the {\it James Webb Space Telescope} ({\it JWST}) 
and the Wide-Field Infrared Survey Telescope (WFIRST) at $z \sim$ 20 \citep[see also][]{
kasen2011,pan12a,wet12b,wet12a,wet12e,hum12}. These explosions, together with other 
Population III (Pop III) SNe \citep{candace2010,wet12c,wet12d,jet13a}, could probe the 
properties of the first stars and galaxies \citep{fsg09,glov12,dw12,fg11}, early cosmological 
reionization and chemical enrichment \citep{whalen2004,mbh03,ss07,bsmith09,ritt12,ss13}, 
and the origins of supermassive black holes \citep{vol12}.
  
Understanding the observational signatures of PPISNe is key to properly identifying them as 
more of them are discovered by the new SN factories such as the Palomar Transient Factory 
\citep[PTF;][]{ptf1}, the Panoramic Survey Telescope and Rapid Response System 
\citep[Pan-STARRS;][]{panstarrs} and the Large Synoptic Survey Telescope \citep[LSST;][]{
lsst}.  High-$z$ SN surveys by {\it JWST} and WFIRST may harvest even greater numbers of 
PPISNe if, as many suspect, the Pop III initial mass function (IMF) is top-heavy.  \citet{
woosley2007} modeled a PPISN in one dimension (1D) with the \KEPLER{} code \citep{kepler,
heger2001} and its light curve with the \STELLA{} code \citep{stella}.  In these simulations, 
{\color{black} a large density spike formed during the collision between the second two pulses 
and the first.}  Rapid variations in the density of the spike due to a radiative instability \citep{
chev82,imam84} imposed large fluctuations in the bolometric luminosity. \citet{wet13d} noted 
similar features in the near infrared (NIR) light curves of the same PPISN modeled with the 
Los Alamos \RAGE{} code. Such spikes would probably not appear in multidimensional flows 
because hydrodynamic instabilities would likely erase the sharp interface between the pulses.  
It is not clear how these processes would change the luminosity of the collision.

Radiation hydrodynamical simulations in at least two dimensions (2D) with adaptive mesh
refinement (AMR) are clearly required to resolve the thin radiating region between the shells 
during the collision and properly model the light curves of PPISNe.  But such simulations push 
the envelope of even state-of-the-art numerical codes.  As a first step to this goal, we have 
performed the first 2D simulations of a PPISN with the \CASTRO{} code with hydrodynamics
but not radiation transport.  In Section 2 we describe our PPISN model and how it is evolved 
in \CASTRO.  In Section 3 we examine the collision of the shells in 2D in detail, and in Section
4 we show how the evolution of the collision in 2D might change the PPISN light curves of
previous calculations.  We conclude in Section 5.

\section{PPISN Model / Numerical Method}

We take as a fiducial case the PPISN examined in \citet{woosley2007}, whose progenitor was 
a solar-metallicity star with an initial main sequence mass of 110 \Ms.  The mass loss rate was 
artificially reduced so that the star had a total mass of 74.6 \Ms\ and a helium core mass of 
49.9 \Ms\ when it died.  The star produces three violent outbursts. The first, P1, ejects most 
of the hydrogen envelope, making a faint Type II supernova and leaving a residual of 50.7 \Ms, 
just a bit more than the helium core itself. After 6.8 yr, the core again contracts and encounters 
the pair instability, twice in rapid succession.  The total mass of the second and third pulses (P2 
and P3) is 5.1 \Ms\ and their kinetic energy is $6 \times 10^{50}$ erg.  P3 collides with P2 at large 
optical depths that are not visible to an external observer.  These combined shells then overtake 
P1 at $\sim$ 10$^{15}$ cm and speeds of a few 1000 km s$^{-1}$.  About 90\% of the energy of 
this collision is radiated away by 10$^{16}$ cm, with peak bolometric luminosities of $\sim 3 
\times 10^{43}\,\erg\,\sec^{-1}$.

\subsection{\CASTRO{}}

\CASTRO{} is a multi-dimensional AMR astrophysical fluid dynamics code \citep{ann2010,
zhang2011}.  It has an unsplit piecewise parabolic method (PPM) hydro scheme \citep{
woodward1984} and multispecies advection.  The equation of state used in our study was 
taken from \citet{timmes2000}, and has contributions from relativistic $e^-e^+$ pairs of 
arbitrary degeneracy, ions, which are treated as an ideal gas, and photons.  Densities, 
velocities,  temperatures and mass fractions from \KEPLER{} were mapped onto a 2D cylindrical grid 
in \CASTRO{} with the conservative scheme of \citet{chen2011, chen2014}, 
which guarantees that quantities such as energy and mass are strictly conserved.  
Because only one quadrant of the star is mapped onto the mesh, outflow 
and reflecting boundary conditions were set on the upper and lower boundaries in $r$ and 
$z$, respectively.  Three levels of adaptive refinement (for up to 64 times greater resolution 
along each axis) were used to resolve the scales on which instabilities form in the flow, and 
the grid refinement criteria are based on gradients of density, velocity, and pressure.  
We use the monopole approximation for self-gravity, in which a 1D gravitational potential is 
constructed from the radial average of the density and 
then applied to gravitational force updates everywhere in the AMR hierarchy.  This 
approximation is very efficient, and well-suited to the nearly spherical symmetry of the star 
and its pulsations. 
 
\section{PPISN Evolution}

\subsection{Fallback}

In principle, partial fallback from one pulse could collide with a subsequent ejection and seed 
the formation of instabilities and mixing before the pulses themselves collide.  We investigate 
this process with a 2D simulation in \CASTRO{} in which we follow fallback from P2 onto P3.  
This simulation must be performed on a much smaller mesh that excludes P1 in order to 
resolve the length scales of nuclear burning, which powers the pulses, and the onset of any 
fluid instabilities near the core.  It is initialized with a \KEPLER{} output at an intermediate 
time from \citet{woosley2007}, $\sim$ 100 sec before P2.  Our \CASTRO{} mesh is 5 $\times$ 
10$^{13}$ cm on a side with $256^2$ zones at the coarsest level.  As in \KEPLER{}, nuclear burning is calculated 
with a 19-isotope network that includes species from hydrogen through \Ni{} \citep{kepler}.  The star is 
evolved until the expulsion of P3.

\begin{figure}[h]
\plottwo{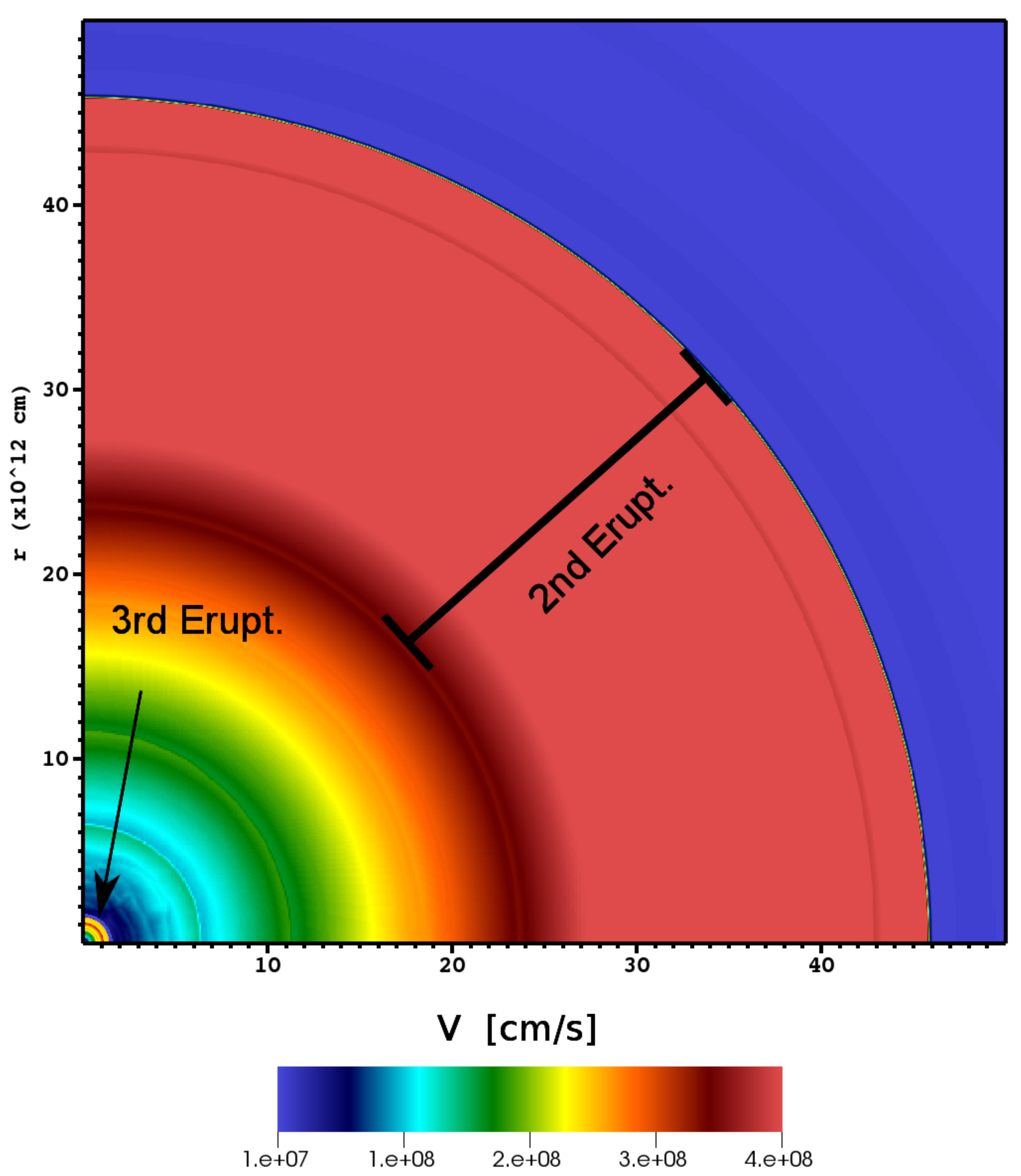}{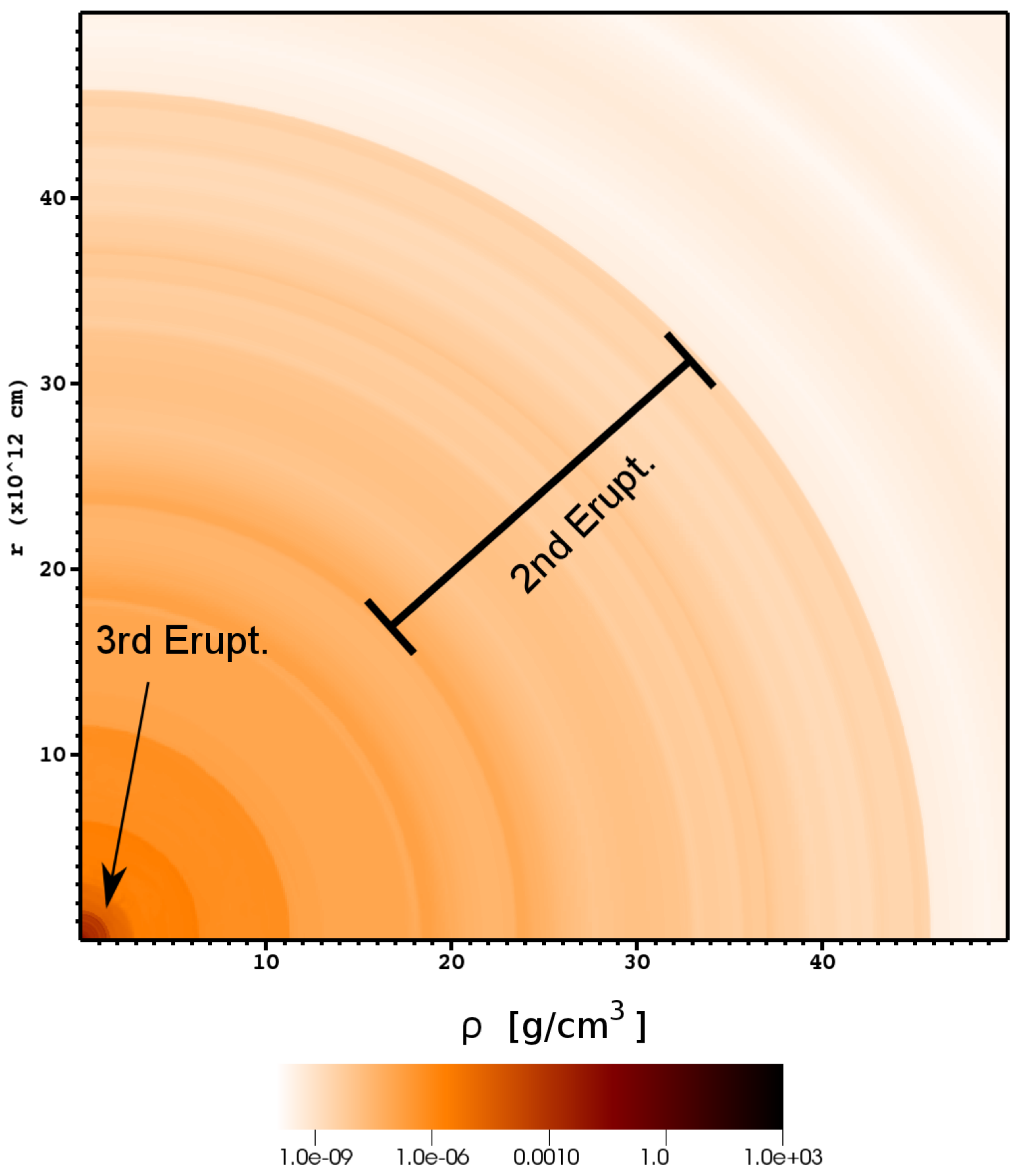}
\caption{ Velocities  and densities at the time P3 is ejected. Minor perturbations in fluid velocities due to fallback from P2 are visible near the surface of the core of the star. The overall density distribution  
is still spherically symmetric. \label{fig:ppsn_p2p3}}
\end{figure}

Gas velocities and densities at the end of this run are shown in Figure~\ref{fig:ppsn_p2p3}.  
Fallback from P2 does trigger mild perturbations in the velocities near the surface of P3, on 
the order of a few percent of the local sound speed, but they do not result in any discernible features in the 
densities.  This simulation demonstrates that launching collision models in \CASTRO{} from 
\KEPLER{} snapshots at later times will not exclude serious dynamical instabilities originating 
from fallback at earlier times.  Modeling perturbations in the gas due to fallback from all three 
pulses at the same time at the required resolution would require a grid that is four orders of 
magnitude larger than this one, so we approximate these features in our collision models by 
seeding the grid with random velocity perturbations of order about $1\,\%$ of local sound 
speed.   
\subsection{First Collision}

\begin{figure}[h]
\begin{center}
\plotone{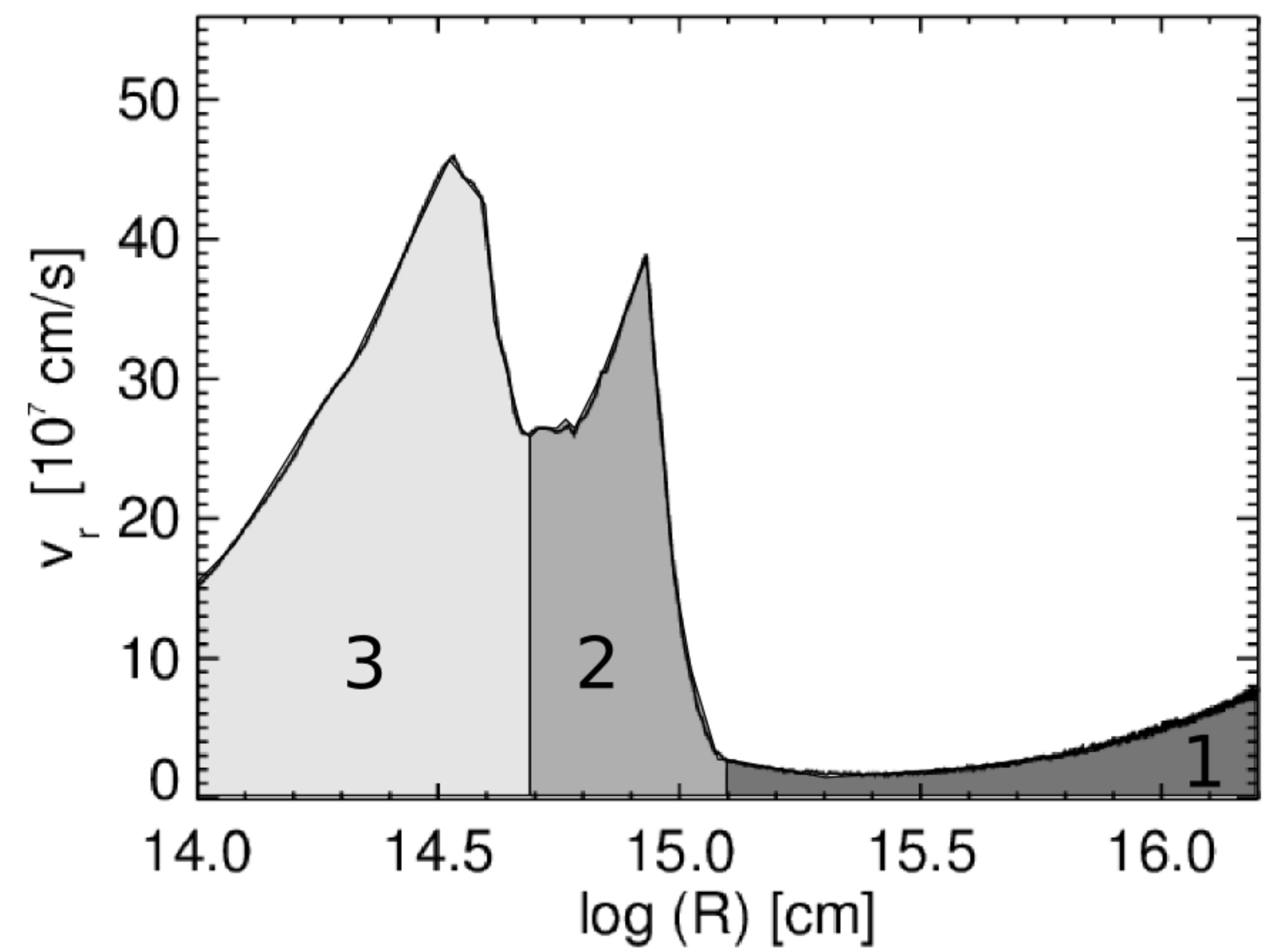} 
\caption{Radial velocities at the beginning of the \CASTRO{} run.  The shaded areas mark 
the ejecta from different eruptions.  P3 has the highest peak velocity, catching up first to P2 
and then eventually to P1. \label{fig:init_v} }
\end{center}
\end{figure}

\begin{figure}[h]
\begin{center}
\includegraphics[width=\columnwidth]{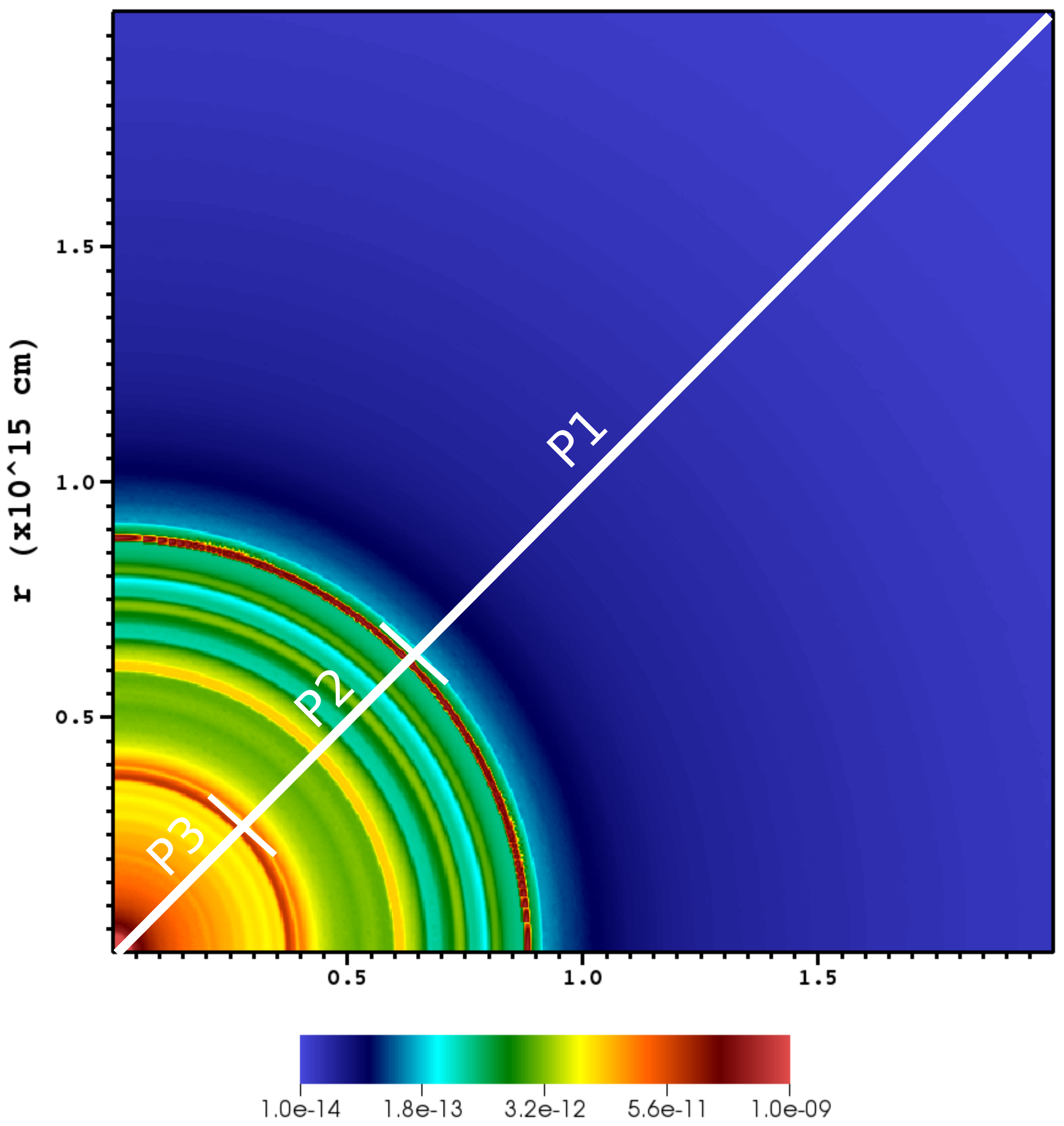} 
\caption{Densities at the beginning of the \CASTRO{} run.  {\color{black} The white tick 
marks mark the boundaries of P1, P2, and P3.}  The finest structure is mostly 
associated with P2.  Note that at the time of launch the mass distribution of the star is 
spherically symmetric.  \label{fig:init_den} }
\end{center}
\end{figure}

To study the collisions between the three pulses, we initialize \CASTRO{} with a \KEPLER{} 
profile taken at a later time, after all three eruptions have occurred and when P1 is at 2 
$\times$ 10$^{16}$ cm, as shown in Figure~\ref{fig:init_v}.  Since \citet{woosley2007} and
\citet{wet13d} both predict that most of the radiation from the collision between shells P2/P3
and P1 is emitted by the time the shock has reached 10$^{16}$ cm, we set the outer 
boundaries of the grid in $r$ and $z$ to be 10$^{16}$ cm.  {\color{black} We do not include all 
of P1 in our simulation because the density beyond $10^{16}$ cm is extremely low and most 
fluid instabilities forming during collisions have become frozen in mass coordinate before 
reaching this boundary.}  With AMR, the effective spatial resolution can be as fine as 10$^{
12}$ cm, which is sufficient for capturing the fine structure of each pulse, as we show in 
Figure~\ref{fig:init_den}) and fluid instabilities in the flow later on.  All three pulses are 
evolved until the collision shock reaches the simulation boundary, about 260 days after the 
launch of the run.

\begin{figure}[h]
\begin{center}
\includegraphics[width=\columnwidth]{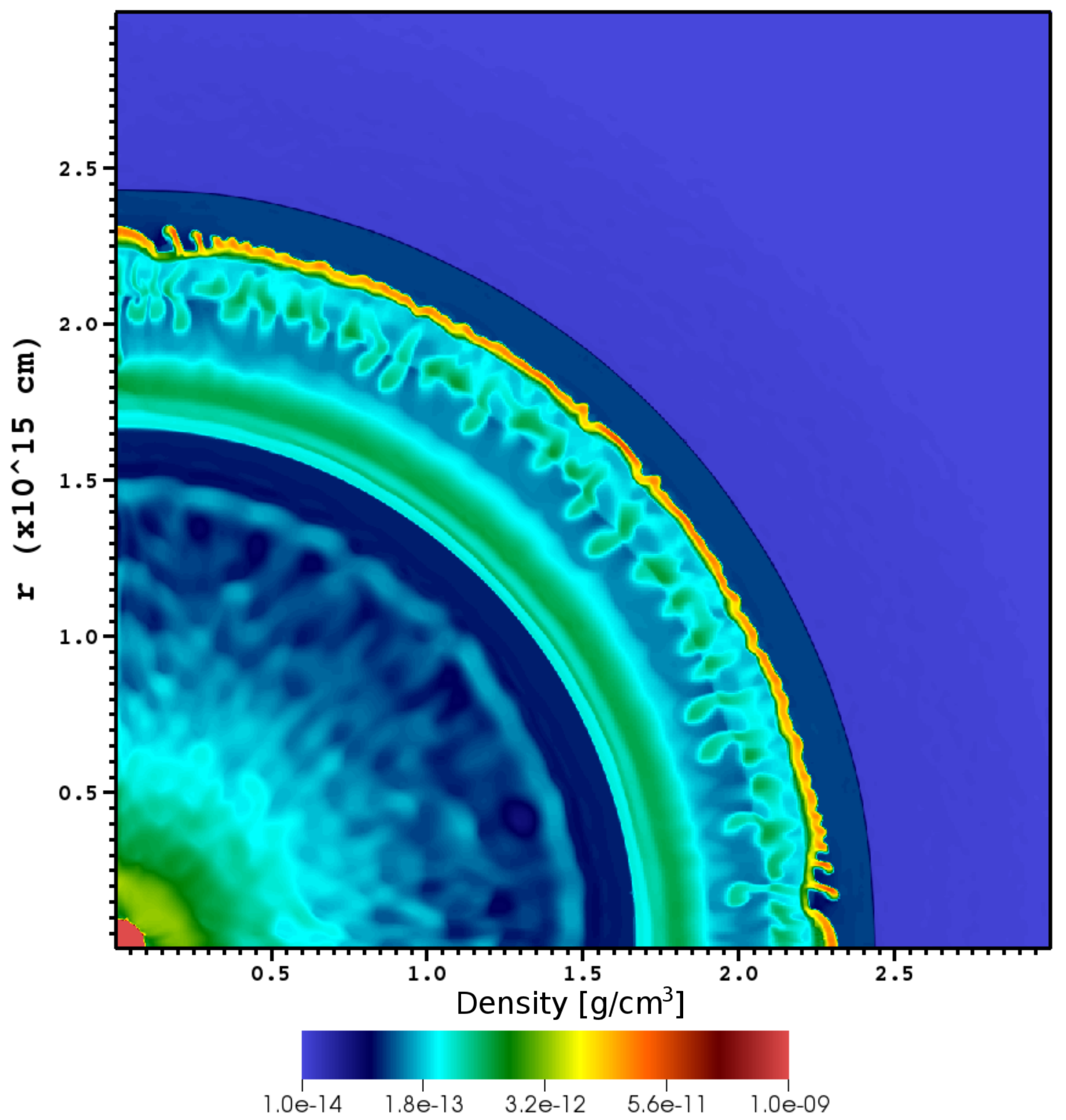} 
\caption{Densities after P3 collides with P2. Mild fluid instabilities have formed and created 
the density fingers at the site of the collision. 
\label{fig:col_den} }
\end{center}
\end{figure}

At the beginning of the simulation, the peak velocities of P2 and P3 are 3.9 $\times$ 
10$^7$ cm s$^{-1}$ and 4.8 $\times$ 10$^7$ cm s$^{-1}$, respectively.  The faster P3 
overtakes P2 within 50 days and has completely merged with it at $r \sim 2.3 \times 10^
{15}\,\cm$.  The contact discontinuity that forms between P3 and P2 upon collision rapidly
destabilizes into the mild, finger-like fluid instabilities that are visible at 2.5 $\times$ 10$^{
15}$ cm in Figure~\ref{fig:col_den}. These fingers are 2 - 3$\times 10^{14}\,\cm$ in size, 
with overdensities of 5 - 10.  The combined pulse has a complex structure with multiple 
velocity peaks, the highest of which is $4.7 \times 10^7$ cm s$^{-1}$.  The bright emission 
from this collision is probably not visible to an external observer because it originates from 
{\color{black} very optically thick regions} that are well below the photosphere at $\sim$ 8 
$\times 10^{15}\,\cm$.  {\color{black} Nevertheless, this radiation may be able affect the 
dynamics of the ejecta or diffuse out of it at later times.} 

\subsection{The Second Collision}

\begin{figure}[h]
\begin{center}
\includegraphics[width=\columnwidth]{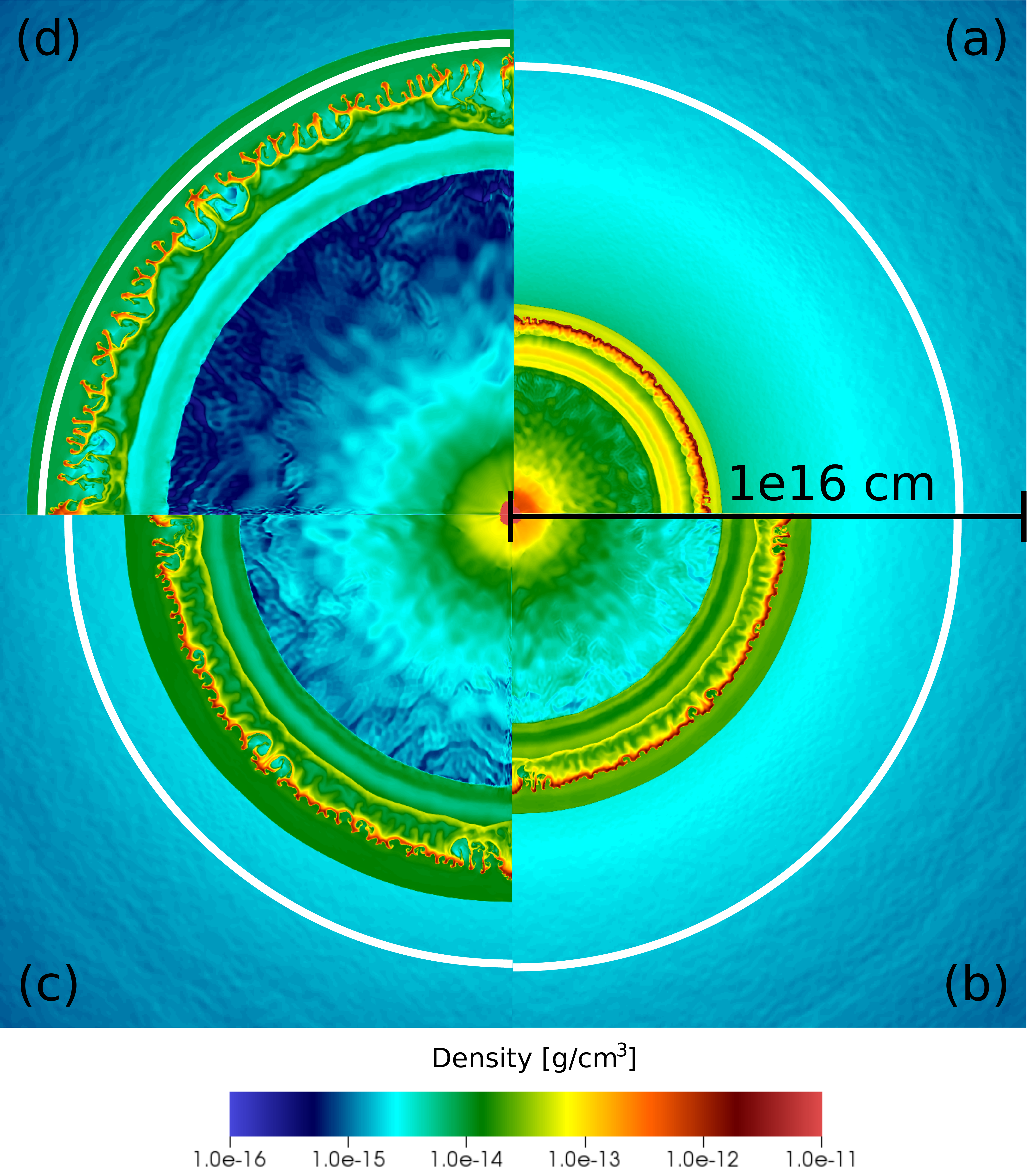} 
\caption{The growth of fluid instabilities and expansion of the photosphere during the second
collision.  Panels (a) - (d) show the instabilities at 97, 149,  205, and 266 days, respectively 
(P2 and P3 have just merged in the first panel).  The white arc in each panel is the 
photosphere. When the shell formed from P2 and P3 
reaches $8\times10^{15}\,\cm$, it becomes visible to an external observer (although most of 
the radiation from their collision has already been emitted).  \label{fig:shell_rev}}
\end{center}
\end{figure}

\begin{figure}[h]
\begin{center}
\includegraphics[width=\columnwidth]{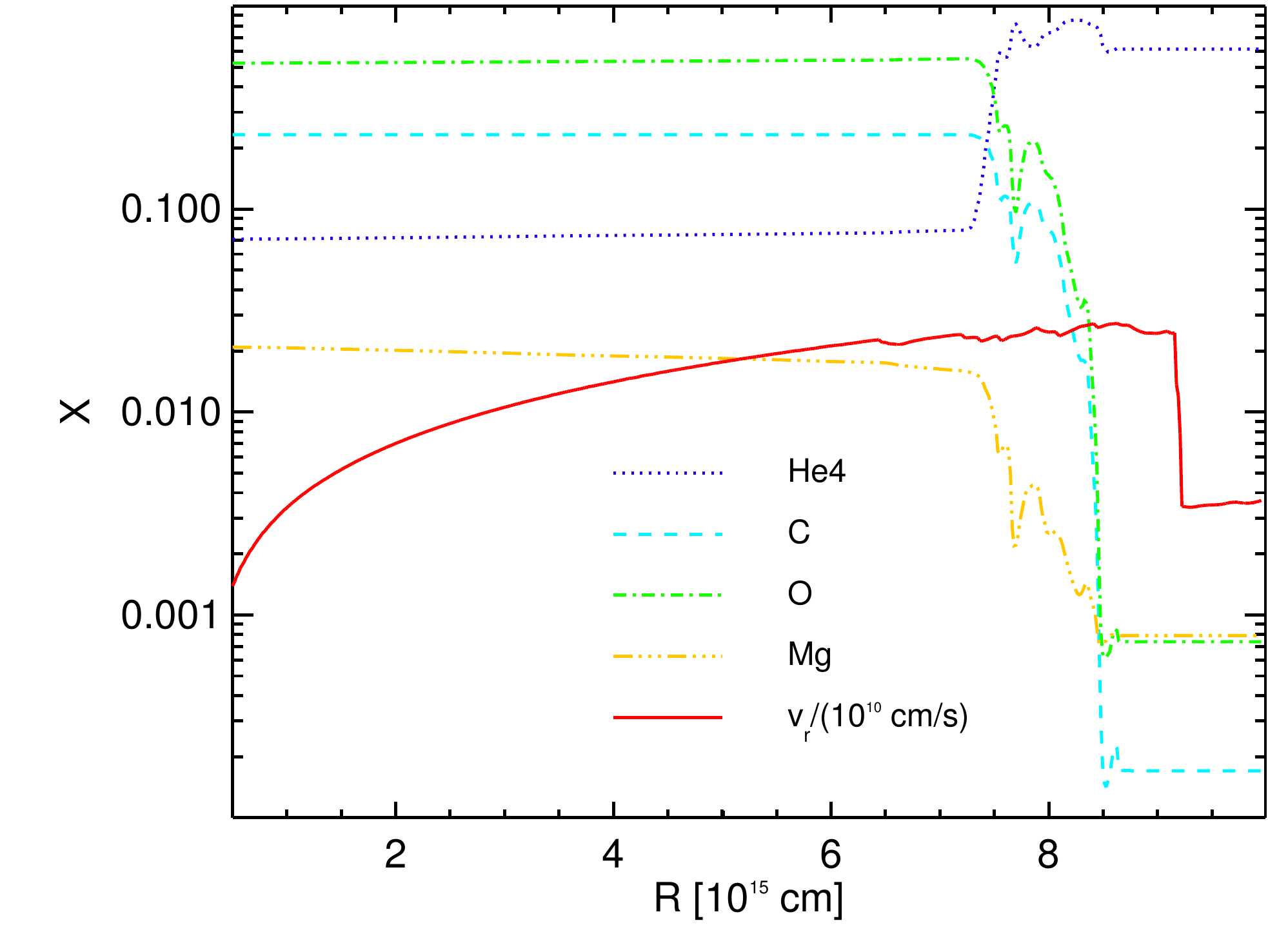} 
\caption{Mass fractions and velocities at 256 days.  The elements that are dredged up are 
mostly \Cx{} and \Ox.  Few elements heavier than \Si{} are ejected by the star, so they may 
not appear in PPISNe. \label{fig:mixing} }
\end{center}
\end{figure}

After P2 and P3 collide, they together begin to overtake P1, which now has a radius 100 
times that of the original star.  Unlike the first collision, the second collision is more like that 
between supernova ejecta and a dense shell expelled by the star prior to its death 
\citep[Type IIn SNe, e.g.,][]{wet12e,moriya2013}.  Although this collision is less violent than 
a Type IIn SN, the instabilities eventually grow to larger amplitudes than in the first collision, 
as we show in Figure~\ref{fig:shell_rev}.  The clumpy structures behind the shock grow in
size as the flow expands, driving more mixing and dredging up heavier elements from 
deeper layers.  We plot angle-averaged mass fractions for \He, \Cx, \Ox, and \Mg{} in 
Figure~\ref{fig:mixing} as the shock approaches the photosphere.  Most mixing happens by 
$\sim$ $10^{16}\,\cm$, and elements heavier than \Si\ remain deeper in the ejecta, so they
would not appear in the spectra.

\begin{figure}[h]
\begin{center}
\includegraphics[width=\columnwidth]{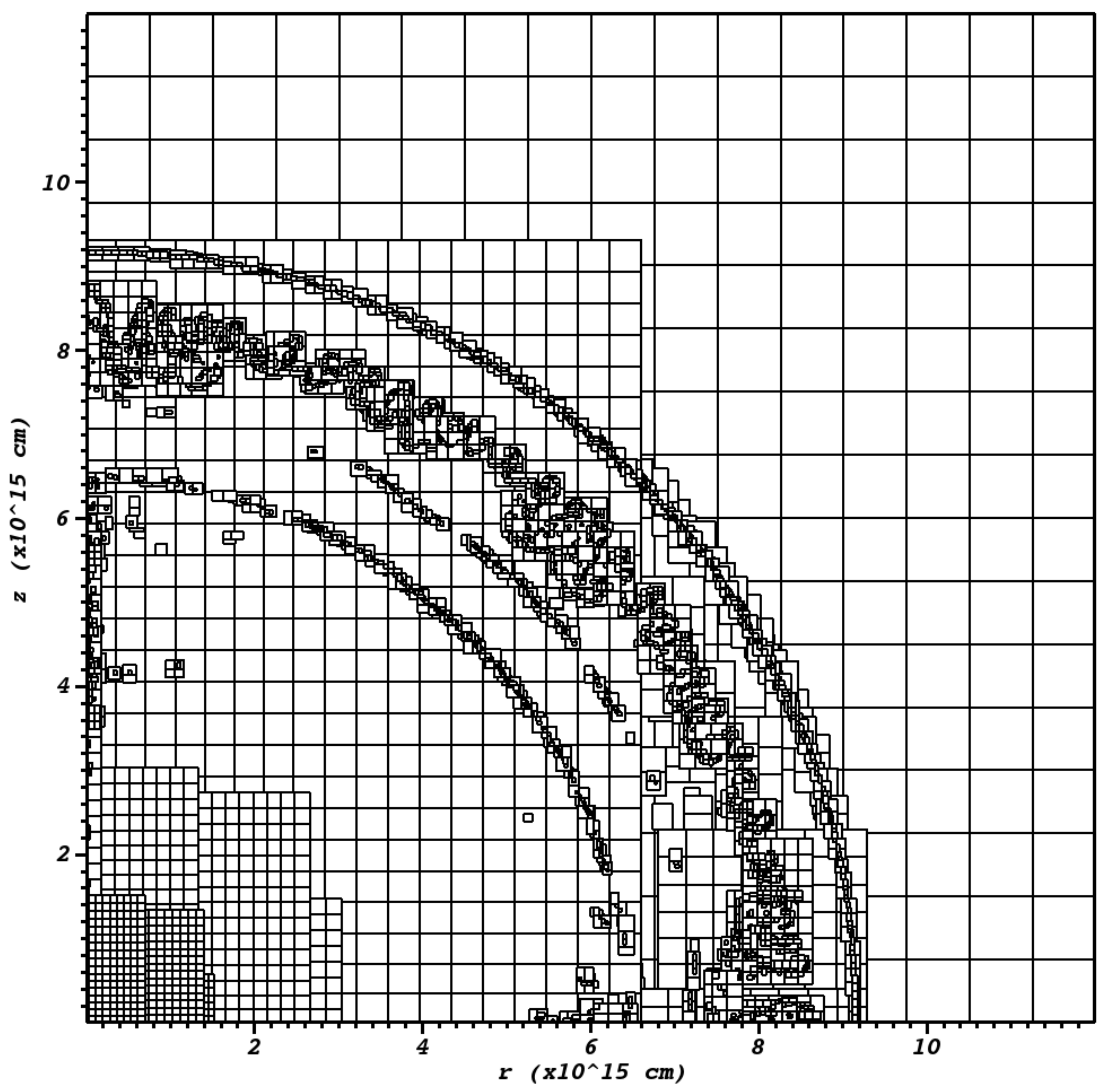} 
\caption{{\color{black} AMR grid structure during the shell collision.  Smaller grids indicate more 
refined zones.  Most of the fine grids are generated to resolve the colliding shells and the core 
of the star.}  \label{fig:grid} }
\end{center}
\end{figure}

\begin{figure}[h]
\begin{center}
\includegraphics[width=\columnwidth]{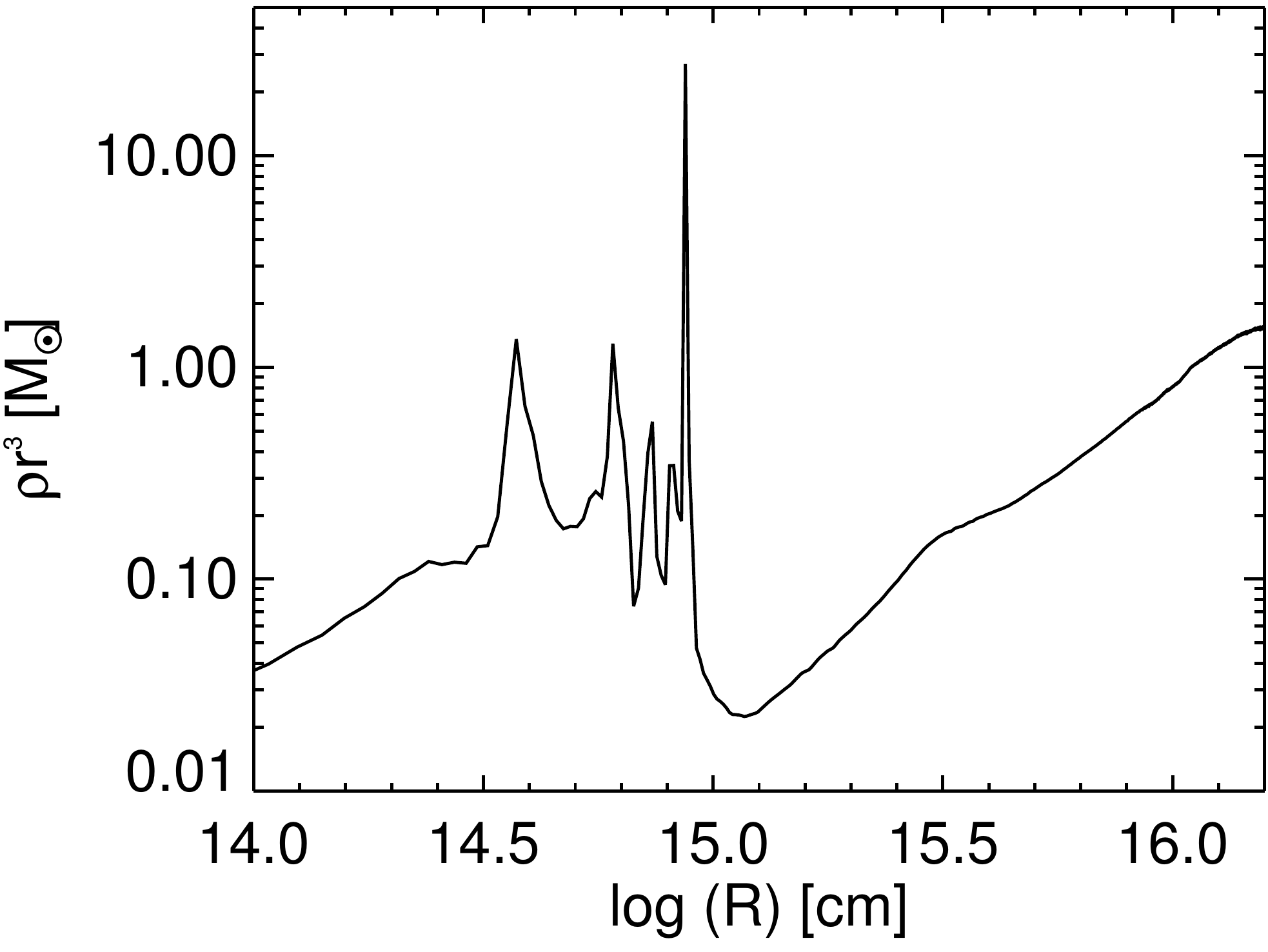} 
\caption{The $\rho\,r^3$ profile at the beginning of the \CASTRO{} run. The ejection of P2 and
P3 create the spiky pattern at $r\,<\,10^{15}\,\cm$.  The ejection and subsequent expansion of 
P1 creates an extended and more homogeneous envelope, which is visible as the steady rise 
of $\rho\,r^3$ at $r \geq 10^{15}\,\cm$. 
\label{fig:rhor3} }
\end{center}
\end{figure}

The formation of dynamical instabilities during the collisions can be understood on analytical
grounds.  Both times, the instabilities appear because the shock decelerates as it plows up 
gas and forms a reverse shock.  For a strong adiabatic shock in a power-law density profile, 
$\rho=Ar^{w}$, the flow becomes self-similar, and the physical quantities that describe it can 
be combined into a single, dimensionless function, $f_{w}(A,E,t)$, where $E$ is the explosion 
energy and $t$ is the time \citep{sedov1959, herant1994}.   The velocity of the shock can 
then be derived from dimensional analysis: 
\begin{equation}
V_{\rm s} = A^{\frac{-1}{(5+w)}}E^{\frac{1}{(w+5)}}t^{\frac{-(w+3)}{5+w}}.
\label{eq:psn_vs}
\end{equation}
For $w>-3$, the shock decelerates as it plows up material. This deceleration is communicated 
to the fluid behind by the shock at the sound speed, and it creates a pressure gradient in the 
direction of the deceleration.  The sound wave generated by the deceleration can steepen this 
pressure gradient, and a reverse shock forms.  If the gas pressure, $P$, and the density, $\rho$,
satisfy the Rayleigh-Taylor (RT) criterion for an fluid \citep{chan1961}
\begin{equation}
\frac{{\rm d} P}{{\rm d}r}\frac{{\rm d}\rho}{{\rm d}r} < 0,
\end{equation}
then instabilities will form.  Whether or not a reverse 
shock forms can be determined by the $\rho r^3$ profile in the surrounding medium. If $\rho 
r^3$ increases with radius, conditions are favorable for the formation of a reverse shock.  We 
plot this quantity for the time at which our \CASTRO{} run is launched in Figure~\ref{fig:rhor3}. 
The peaks and valleys at $r < 10^{15}\,\cm $ were created by the ejection of P2 and P3.  The 
acceleration and deceleration of the shock in this rapidly varying region causes the smaller 
fluid instabilities in the first collision shown in Figure~\ref{fig:col_den}.  The steady increase in
$\rho r^3$ at $r \geq 10^{15}\,\cm$ leading up to P1 allows the instabilities to grow to larger 
amplitudes during the second collision.  When the RT fingers become nonlinear, 
Kelvin-Helmholtz (or shear) instabilities arise and dominate the mixing, which continues until 
the collision shock breaks through P1.   {\color{black} Figure~\ref{fig:grid} shows the AMR grid 
structure when the mixing occurs.  Many fine grids are generated to resolve the clumpy 
structure due to mixing.} 

\subsection{Light Curve}

\begin{figure}[h]
\begin{center}
\includegraphics[width=\columnwidth]{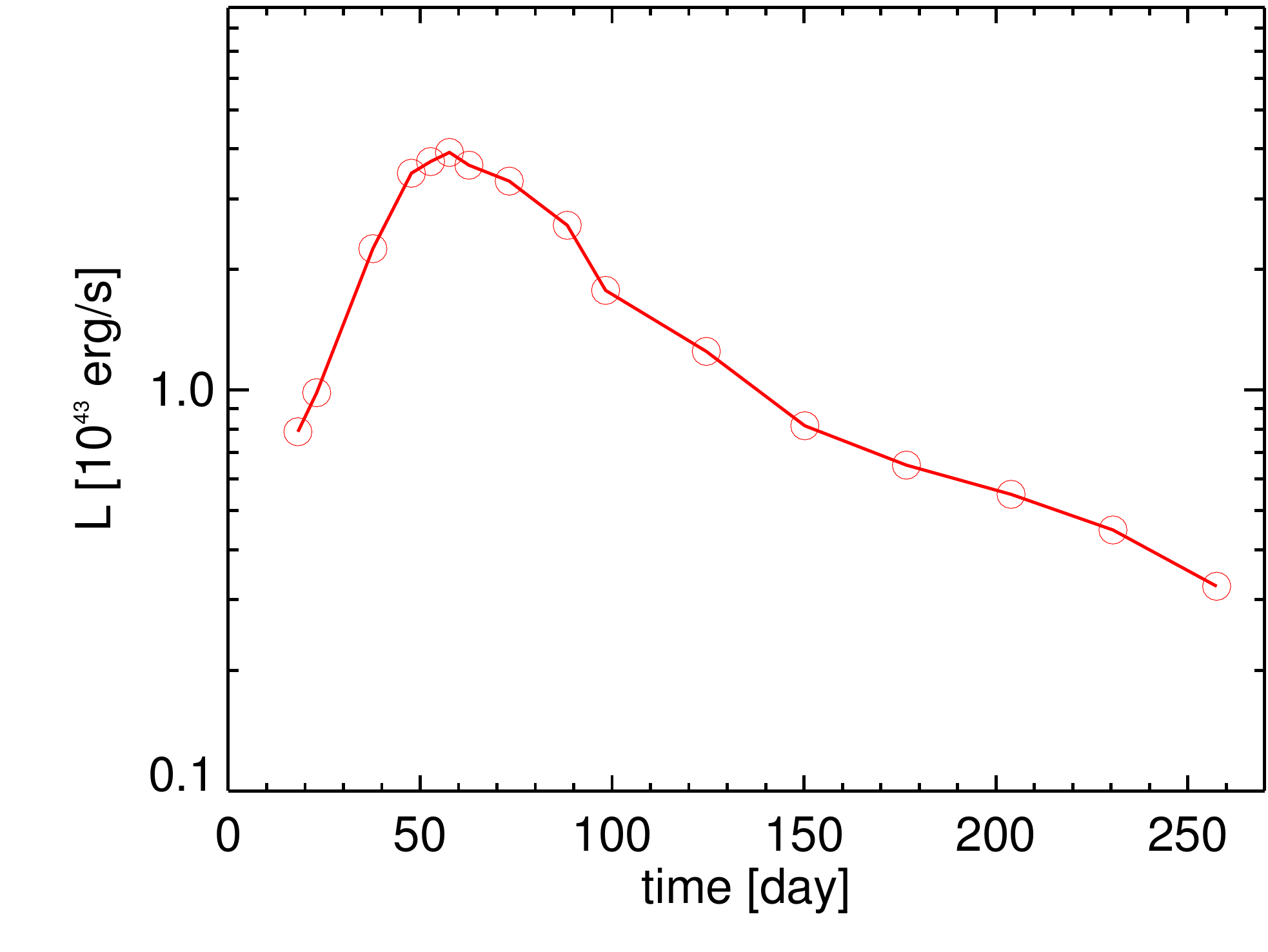} 
\caption{Light curve of shell collision. {\color{black} The red circles represent the 
snapshots from the simulation.} It luminosity peaks at $4\times10^{43}\,\erg\,
\sec^{-1}$ and the duration of emission lasts for 100 days. \label{fig:light}}
\end{center}
\end{figure}

Most of the luminosity from the second collision is gone before mixing ends.  We approximate
the bolometric light curve of this collision by \citep{chevalier2011,chevalier2012}:
\begin{equation}
L_* = 2\pi \rho r^2 v_{s}^3, 
\label{eq:psn_vs}
\end{equation}
where $\rho$ is the gas density just ahead of shock front, $v_{s}$ is the velocity of the shock, 
and $r$ is the position of the shock. Equation~(\ref{eq:psn_vs}) holds if the region in which the 
shock propagates is assumed to be optically thin, and is therefore an upper limit in somewhat 
higher densities. To take into account energy dissipation in the shock by escaping photons, we 
assume that the luminosity of the shock falls as $r^{-2}$ when the shock front is beyond $\sim 
5 \times 10^{15}\,\cm$. {\color{black} We assume that the temperature at the edge of the shock 
falls as $T \propto r^{-1}$ due to thermal radiation, so the bolometric luminosity $L_* \propto 
r^2T^4 \propto r^{-2}.$}  We plot this luminosity in Figure~\ref{fig:light}.  The collision produces 
a very bright transient, with a peak luminosity of $\sim$ $4\times10^{43}\,\erg\,\sec^{-1}$ and a 
duration of $\sim$ 100 days.  {\color{black} The peak luminosity occurs when P2+P3 collide with 
P1.  Its magnitude is calculated with Equation~(\ref{eq:psn_vs}), with $\rho = 1.03\times10^{-14
}\,\gcc$, $r =2.46\times10^{15}\,\cm$, and $v_s = 4.52\times10^{8}$ cm/s.}  The peak 
luminosity and duration are similar to those of the 1D radiation hydrodynamics models of \citet{
woosley2007} and \citet{wet13d}.  But our light curve rises earlier because we assume that the 
environment of the shock is optically thin.  Unlike the 1D light curves, which exhibit artificial 
fluctuations due to a radiative instability, the light curve of our 2D model has a relatively smooth 
peak and no fluctuations.  Our model suggests that radiation hydrodynamical calculations of 
PPISN light curves in 2D will also be smoother than those in 1D because mixing will dampen the 
radiative instability, but that they will have similar luminosities.  {\color{black}
We calculate 2D light curves with Equation (3) with angle-averaged densities and velocities at radii just
ahead of the shock. This light curve is a rough estimate of the luminosity of the shock and does not include radiation from dense clumps due to shell collisions. Although the 2D light curve has similar luminosities  to 
1D models, the light curve rising time and fading tail are still quite different between 1D and 2D.  
Full radiation transport in 2D is needed for more realistic light curves of PPISN models. }

\section{Conclusion}

We have performed the first 2D hydrodynamical simulations of PPISNe with the new \CASTRO{}
code.  We find that mild RT instabilities develop first during the collision of P3 with P2 and then
become stronger as P2 and P3 interact with P1.  Fallback from one pulsation onto another does
not seed strong instabilities but does mildly perturb the flow prior to collisions.  The appearance
of dynamical instabilities and mixing has three consequences for the observational signatures of 
PPISNe that are not captured in 1D models.  

First, the radiative instabilities that cause light curves to fluctuate in 1D models probably do not 
happen in actual explosions.  Mixing between shells dampens the formation of the density spikes 
in 1D models that give rise to these features.  Second, the formation of dense clumps by RT 
fingers could trap photons and alter the luminosity of these events.  We neglect radiative cooling 
by metals and dust in our models so these clumps may catastrophically cool and become much 
denser (and more opaque) in actual collisions.  This in turn could amplify the RT instabilities.  
Finally, since mixing is strongest in the region of the flow from which most of the luminosity of the 
collision originates, it will alter the order in which specific absorption and emission lines appear in 
the spectrum over time.  In particular, our model predicts enhanced spectral lines from \Cx{} and 
\Ox{} at earlier times in the second collision.  The appearance of such lines may prove to be 
powerful diagnostics of mixing in PPISNe.

Since our approximate bolometric luminosities are similar to those calculated with full radiation 
transport in earlier work, future multidimensional calculations will probably not change the recent
result that these explosions will be visible at $z =$ 10 - 20 to {\it JWST} and WFIRST.  But they
do demonstrate that multidimensional radiation transport will be required to model how radiation
is emitted by and escapes from the complex structures formed by dynamical instabilities.  This 
will be key to obtaining more realistic light curves and spectra for PPISNe.  In future papers, we 
will better survey PPISN explosions by examining more progenitor models at low metallicities with 2D 
radiation hydrodynamical calculations \citep{zhang2013}.  PPISNe will soon open new windows
on massive star formation in both the primordial and the local universe.

\acknowledgements 

{\color{black} The authors thank the anonymous referee for reviewing this manuscript and 
providing valuable comments}, and the members 
of CCSE at LBNL for help with \CASTRO{}.  
We also thank Volker Bromm, Dan Kasen, Lars Bildsten, John Bell, and Adam Burrows for many useful 
discussions.  K.C. was supported by an IAU-Gruber Fellowship, a Stanwood Johnston
Fellowship, and a KITP Graduate Fellowship. S.W. acknowledges support by DOE HEP Program 
under contract DE-SC0010676; the National Science Foundation (AST 0909129) and the NASA 
Theory Program  (NNX14AH34G). A.H. was supported by a Future Fellowship 
from the Australian Research Council (ARC FT 120100363). DJW was supported by the 
Baden-W\"{u}rttemberg-Stiftung by contract research via the programme Internationale 
Spitzenforschung II (grant P-LS-SPII/18).  All numerical simulations were performed at the 
University of Minnesota Supercomputing Institute and the National Energy Research Scientific 
Computing Center.  This work has been supported by the DOE grants DE-SC0010676, 
DE-AC02-05CH11231, DE-GF02-87ER40328, DE-FC02-09ER41618 and by NSF grants 
AST-1109394 and PHY02-16783. Work at LANL was done under the auspices of the 
National Nuclear Security Administration of the US Department of Energy at Los Alamos 
National Laboratory under Contract No. DE-AC52-06NA25396.

\newpage

\end{document}